\documentclass[twocolumn]{aastex631}
\usepackage{moresize}
\usepackage{xcolor}
\usepackage{amsmath}
\received{December 23, 2021}
\revised{March 1, 2022}
\accepted{March 10, 2022}

\submitjournal{AJ}

\shorttitle{OCCAM VII: APOGEE DR17 [C/N]-Age Calibration}
\shortauthors{Spoo et al.}
\begin{document}

\title{The Open Cluster Chemical Abundances and Mapping Survey: VII. \\ APOGEE DR17 [C/N]-Age Calibration}

\correspondingauthor{Taylor Spoo}
\email{t.spoo@tcu.edu}

\author[0000-0003-4019-5167]{Taylor Spoo}
\affiliation{Department of Physics and Astronomy, Texas Christian University, TCU Box 298840 \\
Fort Worth, TX 76129, USA (t.spoo, p.frinchaboy, n.myers, j.donor@tcu.edu)}

\author[0000-0002-4818-7885]{Jamie Tayar}
\altaffiliation{NASA Hubble Fellow}
\affiliation{Institute for Astronomy, University of Hawai'i at Mānoa, 2680 Woodlawn Drive, Honolulu, HI 96822, USA}
\affiliation{Department of Astronomy, University of Florida, 211 Bryant Space Science Center, Gainesville, FL 32611, USA}

\author[0000-0002-0740-8346]{Peter M. Frinchaboy}
\affiliation{Department of Physics and Astronomy, Texas Christian University, TCU Box 298840 \\
Fort Worth, TX 76129, USA (t.spoo, p.frinchaboy, n.myers, j.donor@tcu.edu)}

\author{Katia Cunha}
\affiliation{Observatório Nacional, Rua General José Cristino, 77, Rio de Janeiro, RJ 20921-400, Brazil}
\affiliation{Steward Observatory, University of Arizona, 933 North Cherry Avenue, Tucson, AZ 85721-0065, USA}

\author[0000-0001-9738-4829]{Natalie Myers}
\affiliation{Department of Physics and Astronomy, Texas Christian University, TCU Box 298840 \\
Fort Worth, TX 76129, USA (t.spoo, p.frinchaboy, n.myers, j.donor@tcu.edu)}

\author{John Donor}
\affiliation{Department of Physics and Astronomy, Texas Christian University, TCU Box 298840 \\
Fort Worth, TX 76129, USA (t.spoo, p.frinchaboy, n.myers, j.donor@tcu.edu)}

\author{Steven R. Majewski}
\affil{Department of Astronomy, University of Virginia, Charlottesville, VA 22904-4325, USA}

%%%%%%%%%%%%%%%%%%%%%%%%%%%%%%%%%%%%%%%%%%%%%%%%%%%%%%%%%%%%%
%%%%%%%  CURRENT CO-AUTHOR REQUESTS (Alphabetical)   %%%%%%%%
%%%%%%%%%%%%%%%%%%%%%%%%%%%%%%%%%%%%%%%%%%%%%%%%%%%%%%%%%%%%%
\author{Dmitry Bizyaev}
\affiliation{Apache Point Observatory and New Mexico State
University, P.O. Box 59, Sunspot, NM, 88349-0059, USA}
\affiliation{Sternberg Astronomical Institute, Moscow State
University, Moscow, Russia}

\author[0000-0002-1693-2721]{D. A. Garc{\'i}a-Hern{\'a}ndez}
\affiliation{Instituto de Astrof{\'i}sica de Canarias, V{\'i}a L\'actea S/N, 38205 La Laguna, Tenerife, Spain}
\affiliation{Universidad de La Laguna, Departamento de Astrof{\'i}sica, 30206 La Laguna, Tenerife, Spain}

\author[0000-0002-4912-8609]{Henrik J\"onsson}
\affil{Materials Science and Applied Mathematics, Malm\"o University, SE-205 06 Malm\"o, Sweden}

\author[0000-0003-1805-0316]{Richard R. Lane}
\affiliation{Centro de Investigaci{\'o}n en Astronom{\'ia}, Universidad Bernardo O'Higgins, Avenida Viel 1497, Santiago, Chile}
 
\author{Kaike Pan}
\affiliation{Apache Point Observatory and New Mexico State
University, P.O. Box 59, Sunspot, NM, 88349-0059, USA}

\author{Pen{\'e}lope Longa-Pe{\~n}a}
\affiliation{Centro de Astronom{\'i}a, Universidad de Antofagasta, Avenida Angamos 601, Antofagasta 1270300, Chile}

\author[0000-0002-1379-4204]{A. Roman-Lopes} 
\affiliation{Department of Astronomy, Universidad La Serena, La Serena, Chile}

\begin{abstract}

Large scale surveys open the possibility to investigate Galactic evolution both chemically and kinematically, however, reliable stellar ages remain a major challenge. Detailed chemical information provided by high-resolution spectroscopic surveys of the stars 
in clusters can be used as a 
means to calibrate recently developed chemical tools for age-dating field stars.
Using data from the Open Cluster Abundances and Mapping (OCCAM) survey, based on the SDSS/APOGEE-2 survey, we derive a new empirical relationship between open cluster stellar ages and the carbon-to-nitrogen ([C/N]) abundance ratios for evolved stars, primarily those on the red giant branch. {With this calibration, [C/N] can be used a chemical clock for evolved field stars to investigate the formation and evolution of different parts of our Galaxy.}
We explore how mixing effects at different stellar evolutionary phases, like the red clump, affect the derived calibration. 
{We have established the [C/N]-age calibration for APOGEE DR17 giant star abundances to be $\log[Age({\rm yr})]_{\rm DR17} = 10.14 \, (\pm 0.08) + 2.23\,(\pm 0.19) \, {\rm [C/N]}$, usable for $8.62 \leq \log(Age[{\rm yr}]) \leq 9.82$, derived from a uniform sample of 49 clusters observed as part of APOGEE DR17 applicable primarily to metal-rich, thin and thick disk giant stars.} 
This measured [C/N]-age APOGEE DR17 calibration is also shown to be 
consistent with astereoseismic ages derived from Kepler photometry. 

\end{abstract}


\section{Introduction \label{sec:intro}}

To understand our Galaxy's present-day structure, we need to learn how the Milky Way formed and evolved over time. To do this we need to be able to determine the ages of individual stars. However, 
ages are difficult to measure accurately for stars outside of star clusters. 
A common method to determine stellar ages is to compare key parameters, 
whether directly observed (e.g., colors and apparent magnitudes, the latter combined with parallaxes) or 
less directly inferred (e.g., $\log{g}$, $T_{\rm eff}$), to those predicted from stellar evolution models, which allows accurate relative aging of stars, but this strategy is complicated due to degeneracies in these parameters caused by chemical differences. 
Fortunately, several large-scale spectroscopic surveys systematically collecting 
high-resolution 
data, such as {the Apache Point Observatory Galactic Evolution Experiment survey \citep[APOGEE,][]{dr17}, the Gaia-ESO public spectroscopic survey \citep{gaia_eso}, and the GALactic Archaeology with HERMES survey \citep[GALAH,][]{galah},}
are providing detailed and precise chemical abundances for multiple 
chemical elements spanning millions of stars across various Galactic stellar populations, including star clusters.  Apart from providing essential information to break the age-dating degeneracies, these spectroscopic surveys provide additional
key parameters, such as Li-depletion \citep[e.g.,][and references therein]{galahli,lithium} or [C/N] abundance ratios \citep[e.g.,][]{CN_M67,CN_sten,casali_2019}, that are alternative, chemistry-based diagnostics of stellar age.

{A natural ally in calibrating [C/N] ages is asteroseimology, now being generated for large numbers of stars through space missions like CoRoT \citep{corot}, {\em Kepler} \citep{kepler}, and TESS \citep{TESS}. Asteroseismology uses high-precision photometry to measure frequency modes in stars that are dependant on the structure, and therefore mass of the star. Ages for red giant branch (RGB) stars can be derived using asteroseimic photometric measurements from {\em Kepler} and TESS because mass of RGB stars correlates with age. For example, utilizing combined data from {\em Kepler} and the APOGEE survey, the APOKASC catalog \citep{apokasc} has measured masses, and thereby ages, for $\sim 7000$ evolved stars.}

Such masses and ages have been shown previously to correlate with [C/N] \citep{Martig_2016, Ness_2016}. 
However, to obtain their exact absolute calibration is challenging \citep{Gualme_2016, Huber_2017, Zinn_2019}, as is controlling for physical (e.g. mass loss) and systematic differences. Thus, it is worthwhile searching for independent strategies for [C/N]-age calibration.

{Star clusters have long been one of the key testbeds of our understanding of stellar evolution and stellar physics, empirically demonstrating what properties stars of the same chemistry and age should have. In particular, open clusters are useful objects for calibrating the relationship between age and [C/N], because star clusters are the most reliable age-datable tracers, and open clusters also have ages that span the history of the Galactic disk, from currently forming stars up to systems 9-10 Gyr old. Clusters ages can be accurately determined, using the best measured and understood stars in the cluster, by the location of the main sequence turn off (MSTO) on a HR-diagram. }

{ Carbon and nitrogen abundances in evolved stars on the RGB will change due to the first dredge up, where material from regions that have previously undergone nuclear burning are brought to the surface. The changes in the balance of carbon and nitrogen in the surface is dependant on stellar mass \citep{Masseron_Gilmore_2015, Martig_2016, Ness_2016}. With open cluster studies, we can therefore empirically calibrate a relationship between stellar ages, exploiting the precise ages determined by cluster MSTO isochrone-fitting, and precise [C/N] abundances measured for the RGB stars in these systems. }

\citet{casali_2019} has investigated this relationship using a compilation of nearly 40 open clusters using a combined data set from the Gaia-ESO survey and the APOGEE Data Release 14 (DR14) survey.  However, these results relied on a smaller number of clusters studied heterogeneously, which opens the possibility of systematics that could bias the results.

{With the availability of much more APOGEE data via SDSS DR17 --- including the addition of Southern Hemisphere stars and clusters --- we are in a position to improve upon the \citet{casali_2019} effort to calibrate empirically a relationship between [C/N] abundances of RGB stars and their ages using a homogeneous dataset of chemical abundance measurements for open cluster RGB stars deriving entirely from APOGEE spectroscopy.
Using the ages determined by \citet{cg20} for clusters that are in
APOGEE/SDSS DR17 as defined by the Open Cluster Chemical Abundances and Mapping (OCCAM) Survey} (Myers et. al., {\it submitted.}), we calibrate a relationship between [C/N] and stellar age for stars that have experienced the first dredge-up. We separate the RGB and red clump (RC) stars to investigate the effects that stellar evolution may have on the calibration for stellar age. {We also 
compare our results to those from previous work, in particular from the APOKASC astroseismology method for determining stellar ages.  These improvements in the [C/N]-age calibration will help increase the power of large-scale, high resolution stellar spectroscopic surveys like APOGEE to explore Galactic chemodynamical evolution using a reliable means to benchmark ages for vast numbers of stars. }

\section{Data \& Analysis} \label{sec:data_analysis}

\subsection{SDSS/APOGEE Survey} \label{sdss_data}

This study makes use of the Apache Point Observatory Galactic Evolution Experiment surveys \citep[APOGEE-1 \& -2;][Majewski et al., {\it in prep}]{apogee} which are part of the Sloan Digital Sky Survey III \& IV \citep{sdss3,sdss4}, respectively. Data were taken with the 2.5-m Sloan Foundation Telescope \citep{sloan_telescope} at the Apache Point Observatory and the 2.5-m du Pont telescope \citep{du_pont} at the Las Campanas Observatory, using the APOGEE-North and -South spectrographs \citep{apogee_inst}, respectively.  
APOGEE Target selection is described in \citet{zasowski13,zasowski17}, \citet{ap2n_target}, and \citet{ap2s_target}, with open cluster targeting {further} detailed in \citet{frinchaboy_13} and \citet{donor_18}.

The APOGEE spectra were reduced using the APOGEE reduction pipeline \citep{nidever_2015}, and key parameters were
derived automatically via the APOGEE Stellar Parameters and Chemical Abundances Pipeline
\citep[ASPCAP;][]{aspcap}. As the basis of the spectral analysis large grids of synthetic spectra are calculated \citep{zamora_15} using stellar atmospheric models \citep{meszaros_2012} and an updated line list with astrophysically tuned and vetted atomic and molecular data (\citealt{smith21}; see also \citealt{shetrone_2015}). Many aspects of these parts of the data analysis have been updated during the years of the survey and are described in 
\citet{holtzman_2015,holtzman_2018}, \citet{jonsson_2018,jonsson_2020}, \citet{smith21}, and Holtzman et al., ({\em in prep.}). {These continuing analysis updates result in 
the possibility that the} determined stellar parameters and abundances may change for a particular star from one data release to another.
In DR17, ASPCAP reports stellar parameters and chemical abundances for more than 20 elements.
{This includes the abundances of carbon and nitrogen, which can be more challenging to measure from optical spectra.  However, APOGEE is uniquely suited to reliably measuring the C, N, and O abundances, derived through the combined analysis of their contributions to numerous molecular bands of CN, CO, and OH. {These particular molecular lines are fully described in \citet{smith21}.}}

\subsubsection{SDSS/APOGEE Survey DR17}\label{surveyDR17}

In this work, we use data from the latest, and final, data release of the APOGEE-2 survey. The APOGEE-2 Data Release \citep[DR17;][]{dr17} includes {all observations from both APOGEE spectrographs taken from August 2011 to January 2021} and has $\sim$734,000 stars.
A full description of the APOGEE DR17 data quality and parameter limitations will be presented in \citet{dr17} and Holtzman et al., ({\it in prep}). 
{The most significant update in the ASPCAP analysis code for DR17 compared to that used in previous data releases is that the library of}
synthetic spectra were calculated using the Synspec code \citep[e.g.,][]{synspec}.  Synspec allows for the Non-Local Thermodynamic Equilibrium (non-LTE) line formation in a plane parallel geometry. 
{However,} in DR17 only the elements Na, Mg, K and Ca 
were computed in non-LTE \citep[][]{Osorio_2020_NLTE}, while the remaining elements studied by APOGEE, including C and N, were still computed in LTE.
{That said, 
the determined C and N abundances --- and hence [C/N] --- can be different in DR17 from prior data releases,  even in cases where the same spectra are analyzed, because the DR14 and DR16 APOGEE data releases relied on the Turbospectrum code \citep{turbospec} which, while not accommodating non-LTE populations to be specified, does compute atmospheres within a spherical geometry {\citep{Gustafsson2008}}.} {For the interested reader, the choice of the model atmospheres and compositions between the various options is described more fully in \citet{dr17} and Holtzman et al., {\em in prep}.}

\subsubsection{OCCAM Survey Catalog \label{occam}}

This work uses data from the Open Cluster Chemical Analysis and Mapping \citep[OCCAM,][]{frinchaboy_13,cunha_15,donor_18,donor_2020}. 
The DR17 OCCAM data set (Myers et al., {\it submitted}) includes APOGEE data from 153 open clusters 
with a total number of 2061 
stars determined to be likely members.  {From this parent catalog,} Myers et al. select only 94 clusters 
determined to be ``highly reliable” for use in studying Galactic chemical trends, and we refer the reader to  \citet{donor_18,donor_2020} for a full description of member selection and explanation of how the “highly reliable” sample was selected.
We utilize only data from the ``highly reliable” OCCAM DR17 value-added catalog (VAC\footnote{https://www.sdss.org/dr17/data\_access/value-added-catalogs/?vac\_id=open-cluster-chemical-abundances-and-mapping-catalog.})

\subsection{Cluster Member Sample Selection}

To select stars having reliable C and N measurements for our analysis, we used ASPCAP bit-wise flags for removing unreliable C and N abundances, along with a requirement that the spectra {from which they were derived have a}  signal-to-noise ratio (S/N) greater than 70. {The bit-wise flags are fully described in \citet{jonsson_2020} and Holtzman et al, {\em in prep}.}
In addition, we cross-matched our sample with the SB2 Catalog \citep{sb2_cat} to check whether any 
stars in our sample
were double-lined spectroscopic binaries (SB2s), and found none to be SB2s.

We also made a cut to our sample based on OCCAM membership probabilities in radial velocity, proper motion, and metallicity. To provide a more reliable sample, we {selected with a tighter requirement of %\replaced{$<30$\%}
{$>30$\%} membership probability in radial velocity, proper motion, and metallicity.} 

In addition, we 
selected stars that lie within a radius two times larger than the radius containing half the members ($r_{50}$) as determined by \citet{cg20}. {These combined criteria provide a well-vetted and pure sample that compares well to other cluster membership catalogs \citep[e.g.,][]{cg20}.}
Because [C/N] abundances only correlate with age for evolved stars, we applied a surface gravity cut 
to ensure we were obtaining 
evolved stars that have experienced the first dredge-up (FDU). {Based on a combination of stellar models and the APOKASC data, we included only stars with $\log g<3.3$, which should represent the post dredge-up stars and the ages studies here.}
Next, from our uniform sample, we used the APOGEE DR17 RC Catalog \citep[][Bovy et al.{\it in prep}]{bovy_2014_RC} to flag %\textcolor{red}
{likely RC stars within our sample;  
this will allow us to investigate RC mixing effects that could skew the age calibration.} 
{Likely RC stars included in the APOGEE DR17 RC Catalog were selected based on their location within
{color, metallicity, surface gravity and effective temperature space}, with details fully described in \citet{bovy_2014_RC}.}  

The sample selection criteria adopted are listed in Table \ref{tab:cuts}. 
After applying this criteria, our sample is comprised of 75 clusters given in Table \ref{sum_table}, which includes: cluster names, cluster ages as reported by \citet{cg20}, DR17 median [Fe/H] abundances, and median [C/N] abundances {for the member stars with good measurements.} {Median elemental abundance uncertainties were calculated using the standard error of the mean.} 

\begin{deluxetable*}{llr}[ht!]
\tablecaption{{Stellar Data Quality Selection Criteria}\label{tab:cuts}}
\tablewidth{0pt}
\tabletypesize{\normalsize}
\tablehead{
\colhead{Source} & \colhead{Parameter} & \colhead{Selection} 
}
\startdata
\multicolumn{3}{c}{APOGEE Data Quality Cuts} \\\hline
APOGEE/DR17 & STARFLAG - Stellar Parameters &  $!= 16$\\[-0.5ex]
APOGEE/DR17 & ASPCAP C Flag &  $!= 21$\\[-0.5ex]
APOGEE/DR17 & ASPCAP N Flag & $!= 22$ \\[-0.5ex]
APOGEE/DR17 & ASPCAP Flag - Chemistry  &  $!= 23$\\[-0.5ex]
APOGEE/DR17 & VSCATTER (km s$^{-1}$) & $< 1$ \phantom{j}\\[-0.5ex]
APOGEE/DR17 & SNREV & $> 70$ \\
\hline \multicolumn{3}{c}{Cluster Membership Cuts }\\\hline
(Myers et al., {\em submitted}) & RV\_PROB & $> 0.3$ \\[-0.5ex]
(Myers et al., {\em submitted}) & FEH\_PROB & $> 0.3$ \\[-0.5ex]
(Myers et al., {\em submitted}) & PM\_PROB & $> 0.3$ \\[-0.5ex]
\citep{cg20} & $\sqrt{((l_{*} - l_{CG})\cos (b_{CG}))^2 + (b_{*} - b_{CG} )^2 }$ & $< 2 \times$ r50 \\
\hline \multicolumn{3}{c}{Stellar Evolutionary Cuts} \\\hline
APOGEE/DR17 & LOGG & $< 3.3$ %\\[-0.5ex]
\enddata
\tablecomments{The table summarizes the various selection criteria applied to the stars within clusters common between \citet{cg20} and the OCCAM survey VAC (Myers et al., {\it submitted}) to ensure that only cluster stars whose properties are well measured and for which the [C/N] ratios should be correlated with age. {STARFLAG and ASPCAP bit-wise flags are described in: https://www.sdss.org/dr17/irspec/apogee-bitmasks/ .}}
\end{deluxetable*}

\begin{deluxetable}{lrrrr}[h!]
 \tablecaption{Mean [C/N] DR17 Abundance - Full Sample\label{sum_table}}
 \tablewidth{0pt}
 \tabletypesize{\ssmall}
 \tablehead{
 \colhead{Cluster\tablenotemark{a}} & 
 \colhead{$\log$(Age)\tablenotemark{b}} &  
 \colhead{[Fe/H]} &
 \colhead{[C/N]} &
 \colhead{DR17}\\[-1.5ex]
 \colhead{name} &
 \colhead{(yr)} &
 \colhead{(dex)} & 
 \colhead{(dex)} & 
 \colhead{Memb.} 
 }
 \startdata
{\bf Berkeley 17}  & {\bf 9.86} & {\bf $-$0.183} & {\bf $-0.164 \pm 0.010$} & {\bf 8}\\ [-1ex]
{\bf Berkeley 18 } & {\bf 9.64} & {\bf $-$0.385} & {\bf $-0.332 \pm 0.015$} & {\bf 18} \\ [-1ex]
     Berkeley 19   &      9.34  &      $-$0.361  &      $-0.517 \pm 0.000$  & 1\\ [-1ex]
{\bf Berkeley 2  } & {\bf 8.77} & {\bf $-$0.208} & {\bf $-0.366 \pm 0.048$} & {\bf 6} \\ [-1ex]
     Berkeley 20   &      9.68  &      $-$0.432  &      $-0.372 \pm 0.000$  & 1\\ [-1ex]
{\bf Berkeley 21 } & {\bf 9.33} & {\bf $-$0.269} & {\bf $-0.355 \pm 0.006$} & {\bf 3} \\ [-1ex]
{\bf Berkeley 29 } & {\bf 9.49} & {\bf $-$0.527} & {\bf $-0.280 \pm 0.009$} & {\bf 2} \\ [-1ex]
{\bf Berkeley 31 } & {\bf 9.45} & {\bf $-$0.426} & {\bf $-0.278 \pm 0.001$} & {\bf 2} \\ [-1ex]
     Berkeley 33   &      8.37  &      $-$0.243  &      $-0.560 \pm 0.000$  & 1\\ [-1ex]
{\bf Berkeley 53 } & {\bf 8.99} & {\bf $-$0.121} & {\bf $-0.517 \pm 0.025$} & {\bf 6} \\ [-1ex]
{\bf Berkeley 53 } & {\bf 8.99} & {\bf $-$0.121} & {\bf $-0.517 \pm 0.025$} & {\bf 6} \\ [-1ex]
{\bf Berkeley 66 } & {\bf 9.49} & {\bf $-$0.215} & {\bf $-0.302 \pm 0.008$} & {\bf 5} \\ [-1ex]
{\bf Berkeley 71 } & {\bf 8.94} & {\bf $-$0.232} & {\bf $-0.519 \pm 0.029$} & {\bf 5} \\ [-1ex]
{\bf Berkeley 75 } & {\bf 9.23} & {\bf $-$0.412} & {\bf $-0.504 \pm 0.039$} & {\bf 3} \\ [-1ex]
{\bf Berkeley 85 } & {\bf 8.62} & {\bf $+$0.064} & {\bf $-0.420 \pm 0.009$} & {\bf 10} \\ [-1ex]
     Berkeley 9    &      9.14  &      $-$0.180  &      $-0.624 \pm 0.000$  & 1\\ [-1ex]
     Berkeley 91   &      8.80  &      $+$0.070  &      $-0.327 \pm 0.000$  & 1\\ [-1ex]
{\bf Berkeley 98 } & {\bf 9.39} & {\bf $-$0.044} & {\bf $-0.376 \pm 0.043$} & {\bf 5} \\ [-1ex]
{\bf Czernik 20  } & {\bf 9.22} & {\bf $-$0.177} & {\bf $-0.400 \pm 0.049$} & {\bf 4} \\ [-1ex]
{\bf Czernik 21  } & {\bf 9.41} & {\bf $-$0.326} & {\bf $-0.280 \pm 0.008$} & {\bf 2} \\ [-1ex]
     Czernik 23    &      8.43  &      $-$0.329  &      $-0.571 \pm 0.000$  & 1\\ [-1ex]
{\bf Czernik 30  } & {\bf 9.46} & {\bf $-$0.387} & {\bf $-0.348 \pm 0.017$} & {\bf 2} \\ [-1ex]
{\bf ESO 211 03  } & {\bf 9.11} & {\bf $+$0.107} & {\bf $-0.447 \pm 0.031$} & {\bf 4} \\ [-1ex]
{\bf FSR 0494    } & {\bf 8.95} & {\bf $-$0.021} & {\bf $-0.515 \pm 0.015$} & {\bf 5} \\ [-1ex]
     FSR 0496      &      9.31  &      $-$0.137  &      $-0.429 \pm 0.000$  & 1\\ [-1ex]
     FSR 0667      &      8.85  &      $-$0.034  &      $-0.544 \pm 0.000$  & 1\\ [-1ex]
     FSR 0716      &      8.94  &      $-$0.385  &      $-0.398 \pm 0.000$  & 1\\ [-1ex]
{\bf FSR 0937    } & {\bf 9.08} & {\bf $-$0.371} & {\bf $-0.439 \pm 0.032$} & {\bf 2} \\ [-1ex]
     FSR 1113      &      9.44  &      $-$0.347  &      $-0.237 \pm 0.000$  & 1\\ [-1ex]
     Haffner 4     &      8.66  &      $-$0.182  &      $-0.482 \pm 0.000$  &   1 \\ [-1ex]
     IC 1369       &      8.46  &      $-$0.112  &      $-0.560 \pm 0.024$  & 3\\ [-1ex]
{\bf IC 166      } & {\bf 9.12} & {\bf $-$0.095} & {\bf $-0.488 \pm 0.037$} & {\bf 13} \\ [-1ex]
     King 15       &      8.47  &      $-$0.128  &      $-0.457 \pm 0.000$  & 1\\ [-1ex]
     King 2        &      9.61  &      $-$0.184  &      $-0.292 \pm 0.000$  & 1\\ [-1ex]
{\bf King 5      } & {\bf 9.01} & {\bf $-$0.159} & {\bf $-0.464 \pm 0.013$} & {\bf 4} \\ [-1ex]
     King 7        &      8.35  &      $-$0.162  &      $-0.366 \pm 0.073$  & 7\\ [-1ex]
     King 8        &      8.92  &      $-$0.239  &      $-0.469 \pm 0.000$  & 1\\ [-1ex]
{\bf Melotte 71  } & {\bf 8.99} & {\bf $-$0.151} & {\bf $-0.478 \pm 0.020$} & {\bf 3} \\ [-1ex]
{\bf NGC 1193    } & {\bf 9.71} & {\bf $-$0.345} & {\bf $-0.255 \pm 0.026$} & {\bf 3} \\ [-1ex]
{\bf NGC 1245    } & {\bf 9.08} & {\bf $-$0.089} & {\bf $-0.480 \pm 0.017$} & {\bf 25} \\ [-1ex]
     NGC 136       &       8.41 &      $-$0.266  &      $-0.457 \pm 0.000$  & 1\\ [-1ex]
     NGC 1664      &       8.71 &      $-$0.060  &      $-0.564 \pm 0.000$  & 1\\ [-1ex]
{\bf NGC 1798    } & {\bf 9.22} & {\bf $-$0.279} & {\bf $-0.429 \pm 0.017$} & {\bf 10} \\ [-1ex]
{\bf NGC 1817    } & {\bf 9.05} & {\bf $-$0.157} & {\bf $-0.402 \pm 0.045$} & {\bf 5} \\ [-1ex]
     NGC 1857      &       8.40 &      $-$0.176  &      $-0.414 \pm 0.013$  & 2\\ [-1ex]
{\bf NGC 188     } & {\bf 9.85} & {\bf $+$0.064} & {\bf $-0.244 \pm 0.010$} & {\bf 32} \\ [-1ex]
{\bf NGC 1907    } & {\bf 8.77 }& {\bf $-$0.126 }& {\bf $-0.506 \pm 0.046$} & {\bf 3} \\ [-1ex]
{\bf NGC 1912    } & {\bf 8.47} & {\bf $-$0.169} & {\bf $-0.435 \pm 0.016$} & {\bf 2} \\ [-1ex]
{\bf NGC 2158    } & {\bf 9.19} & {\bf $-$0.246} & {\bf $-0.342 \pm 0.013$} & {\bf 41} \\ [-1ex]
{\bf NGC 2204    } & {\bf 9.32} & {\bf $-$0.280} & {\bf $-0.423 \pm 0.021$} & {\bf 20} \\ [-1ex]
{\bf NGC 2243    } & {\bf 9.64} & {\bf $-$0.458} & {\bf $-0.338 \pm 0.037$} & {\bf 6} \\ [-1ex]
{\bf NGC 2304    } & {\bf 8.96} & {\bf $-$0.141} & {\bf $-0.400 \pm 0.009$} & {\bf 2} \\ [-1ex]
{\bf NGC 2324    } & {\bf 8.73} & {\bf $-$0.221} & {\bf $-0.517 \pm 0.037$} & {\bf 4} \\ [-1ex]
{\bf NGC 2420    } & {\bf 9.24} & {\bf $-$0.205} & {\bf $-0.337 \pm 0.012$} & {\bf 16} \\ [-1ex]
{\bf NGC 2447}     & {\bf 8.76} & {\bf $-$0.114} & {\bf $-0.548 \pm 0.016$} & {\bf 3}\\ [-1ex]
     NGC 2479      &       8.99 &      $-$0.049  &      $-0.436 \pm 0.000$  & 1   \\ [-1ex]
{\bf NGC 2682    } & {\bf 9.63} & {\bf $+$0.000} & {\bf $-0.368 \pm 0.011$} & {\bf 19} \\ [-1ex]
{\bf NGC 4337    } & {\bf 9.16} & {\bf $+$0.222} & {\bf $-0.491 \pm 0.012$} & {\bf 6} \\ [-1ex]
     NGC 6705      &       8.49 &      $+$0.096  &      $-0.440 \pm 0.016$  & 12\\ [-1ex]
{\bf NGC 6791    } & {\bf 9.80} & {\bf $+$0.313} & {\bf $-0.165 \pm 0.010$} & {\bf 41} \\ [-1ex]
{\bf NGC 6811    } & {\bf 9.03} & {\bf $-$0.051} & {\bf $-0.521 \pm 0.012$} & {\bf 6} \\ [-1ex]
{\bf NGC 6819    } & {\bf 9.35} & {\bf $+$0.040} & {\bf $-0.352 \pm 0.019$} & {\bf 42} \\ [-1ex]
{\bf NGC 6866    } & {\bf 8.81} & {\bf $+$0.010} & {\bf $-0.565 \pm 0.030$} & {\bf 2} \\ [-1ex]
     NGC 7062      &       8.63 &      $-$0.023  &      $-0.569 \pm 0.000$  & 1\\ [-1ex]
{\bf NGC 752     } & {\bf 9.07} & {\bf $-$0.064} & {\bf $-0.442 \pm 0.010$} & {\bf 5} \\ [-1ex]
{\bf NGC 7789    } & {\bf 9.19} & {\bf $-$0.032} & {\bf $-0.421 \pm 0.011$} & {\bf 65} \\ [-1ex]
{\bf Ruprecht 147} & {\bf 9.48} & {\bf $+$0.116} & {\bf $-0.395 \pm 0.030$} & {\bf 3} \\ [-1ex]
     Ruprecht 82   &       8.66 &      $-$0.037  &      $-0.702 \pm 0.000$  & 1\\ [-1ex]
     Teutsch 10    &       8.79 &      $-$0.350  &      $-0.638 \pm 0.000$  & 1\\ [-1ex]
{\bf Teutsch 12  } & {\bf 8.92} & {\bf $-$0.200} & {\bf $-0.569 \pm 0.031$} & {\bf 4} \\ [-1ex]
{\bf Teutsch 51  } & {\bf 8.83} & {\bf $-$0.332} & {\bf $-0.524 \pm 0.011$} & {\bf 3} \\ [-1ex]
     Teutsch 84    &      9.02  &      $+$0.200 &       $-0.506 \pm 0.000$  & 1\\ [-1ex]
{\bf Tombaugh 2  } & {\bf 9.21} & {\bf $-$0.371} & {\bf $-0.605 \pm 0.070$} & {\bf 4} \\ [-1ex]
{\bf Trumpler 20 } & {\bf 9.27} & {\bf $+$0.105} & {\bf $-0.509 \pm 0.017$} & {\bf 25} \\ [-1ex]
{\bf Trumpler 5  } & {\bf 9.63} & {\bf $-$0.449} & {\bf $-0.274 \pm 0.018$} & {\bf 8} \\
\enddata
 \tablenotetext{a}{ {\bf Bold} denotes clusters in the final calibration sample. }
 \tablenotetext{b}{ All ages come from \citet{cg20} and the $\log(Age)$ errors are equal to 0.1.}
 \end{deluxetable}         

 \begin{deluxetable*}{llrrrrc}[ht!]
\tablecaption{Full calibration cluster sample stellar data from APOGEE DR17. \label{tab:full_sample_stars}}
\tablewidth{0pt}
\tabletypesize{\normalsize}
\tablehead{
\colhead{Cluster} & 
\colhead{2MASS ID} &
\colhead{$T_{eff}$} &
\colhead{$\log(g)$} &
\colhead{[Fe/H]} & 
\colhead{[C/N]} &
\colhead{RC? \tablenotemark{a}}\\[-0.5ex]
\colhead{name} &
\colhead{} &
\colhead{(K)} &
\colhead{(dex)} & %$[cm/s^2]$ for mrt
\colhead{(dex)} & 
\colhead{(dex)} & 
\colhead{label} 
}
\startdata
Berkeley 17 & 2M05195385$+$3035095 & 4665 $\pm$ \phn9 & 2.60 $\pm$ 0.03 & $-$0.12 $\pm$ 0.01 & $-$0.23 $\pm$ 0.02 & N \\
Berkeley 17 & 2M05202118$+$3035544 & 4799 $\pm$ \phn9 & 2.47 $\pm$ 0.02 & $-$0.15 $\pm$ 0.01 & $-$0.16 $\pm$ 0.02 & N \\
Berkeley 17 & 2M05202905$+$3032414 & 4783 $\pm$    13 & 2.84 $\pm$ 0.03 & $-$0.20 $\pm$ 0.01 & $-$0.18 $\pm$ 0.03 & N \\
Berkeley 17 & 2M05203121$+$3035067 & 4820 $\pm$ \phn9 & 2.49 $\pm$ 0.02 & $-$0.19 $\pm$ 0.01 & $-$0.14 $\pm$ 0.02 & N \\
Berkeley 17 & 2M05203650$+$3030351 & 4444 $\pm$ \phn6 & 1.98 $\pm$ 0.02 & $-$0.17 $\pm$ 0.01 & $-$0.15 $\pm$ 0.02 & N \\
Berkeley 17 & 2M05203799$+$3034414 & 4307 $\pm$ \phn6 & 1.93 $\pm$ 0.02 & $-$0.18 $\pm$ 0.01 & $-$0.16 $\pm$ 0.02 & N \\
Berkeley 17 & 2M05204143$+$3036042 & 4824 $\pm$ \phn9 & 2.49 $\pm$ 0.03 & $-$0.22 $\pm$ 0.01 & $-$0.14 $\pm$ 0.02 & Y \\
Berkeley 17 & 2M05204488$+$3038020 & 4807 $\pm$ \phn8 & 2.43 $\pm$ 0.02 & $-$0.21 $\pm$ 0.01 & $-$0.21 $\pm$ 0.02 & Y \\ \hline
Berkeley 18 & 2M05211671$+$4533170 & 4220 $\pm$ \phn6 & 1.43 $\pm$ 0.02 & $-$0.40 $\pm$ 0.01 & $-$0.32 $\pm$ 0.02 & N \\
Berkeley 18 & 2M05214927$+$4525225 & 5126 $\pm$    19 & 2.65 $\pm$ 0.04 & $-$0.33 $\pm$ 0.01 & $-$0.26 $\pm$ 0.05 & Y \\
Berkeley 18 & 2M05215476$+$4526226 & 4309 $\pm$ \phn6 & 1.56 $\pm$ 0.03 & $-$0.40 $\pm$ 0.01 & $-$0.35 $\pm$ 0.02 & N \\
\multicolumn{7}{c}{\nodata}\\
\enddata
\tablenotetext{a}{ `Y' is an identified red clump star using the APOGEE DR17 red clump
      value-added catalog (Bovy et al., 2014). \\ \phn `N' denotes stars not
	  identified as part of the catalog and considered red giant branch
	  stars.}
\tablecomments{This table is available in its entirety in machine-readable form in the online journal. A portion is shown here for guidance regarding its form and content.}
\end{deluxetable*}

\subsection{Systematics between DR14, DR16, \& DR17 \label{system}}

During the course of the APOGEE survey, the APOGEE team has made improvements to the automated analysis pipeline.
For each APOGEE data release there have been changes and improvements to, for example, the atomic and molecular line lists, stellar atmosphere models, 
implementation of the code, etc, as discussed in \S \ref{sdss_data}. 
{{In order to} compare our work with results from other studies that used previous APOGEE data releases \citep[e.g.,][ which partial used DR14 data]{casali_2019}, we briefly investigate 
systematic differences between the carbon and nitrogen abundance results derived from the different APOGEE data releases.} 

The APOGEE-2 Data Release 14 \citep[DR14;][]{dr14} included observations from August 2011 to July 2016 and has $\sim 263,000$ stars.  DR14 APOGEE data includes data from the APOGEE 1 and first two years of the APOGEE-2 North survey, all taken from the Apache Point Observatory.  A full description of the APOGEE DR14 data quality and parameter limitations is presented in \citet[][]{holtzman_2018, jonsson_2018}.
The APOGEE-2 Data Release 16 
\citep[DR16;][]{dr16} included observations from August 2011 to August 2018 and has $\sim 473,000$ stars.  DR16 APOGEE data included data from the APOGEE 1 and first four years of the APOGEE-2 North survey taken from the Apache Point Observatory, plus the first year of APOGEE-2 South data taken from the Las Campanas Observatory.  A full description of the APOGEE DR16 data quality and parameter limitations is presented in and \citet{jonsson_2020}.

{We find a global offset of [Fe/H] between DR14 and DR17 as well as for DR16 and DR17 to be 0.07 dex and 0.02 dex, respectively. This investigation for differences between previous releases for metallicity ([Fe/H]), $\alpha$ elements (O, Mg, Si, S, Ca, Ti), iron-peak elements (V, Cr, Mn, Co, Ni), and other elements (Na, Al, K) is discussed in detail in \citet{donor_2020} and Myers et al., ({\it submitted}).}

\begin{figure*}
    \centering
    \epsscale{1.18}\plotone{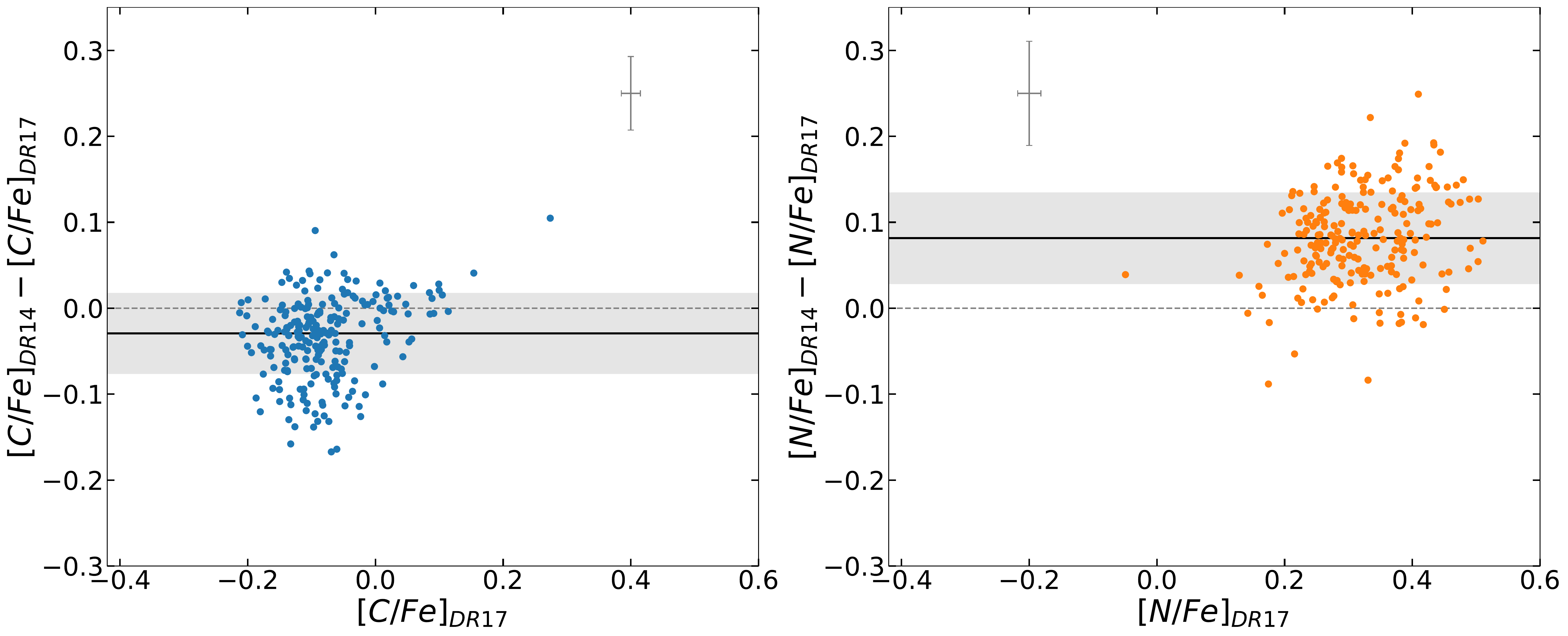}%Figures/compared_DR14DR17_CN_abundances_individualstars_wShadedRegion.pdf}
    \plotone{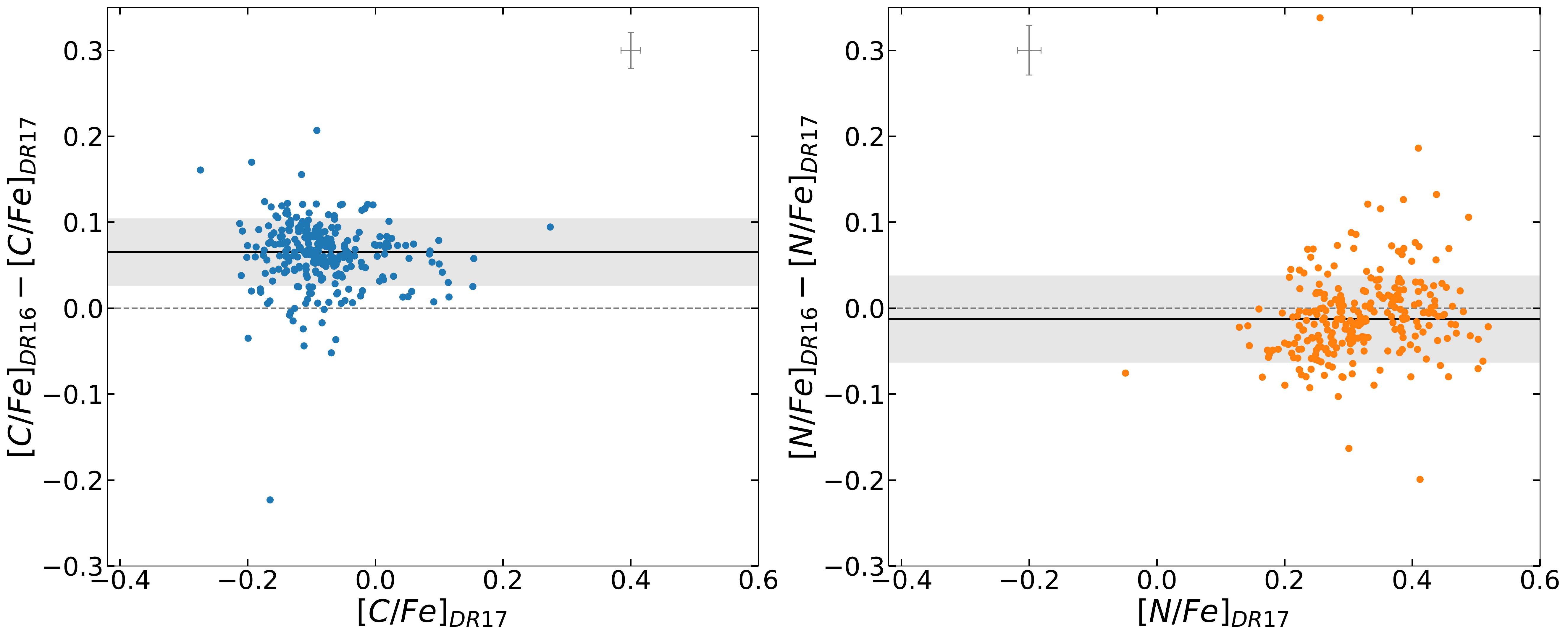}%compared_DR16DR17_CN_abundances_individualstars_wShadedRegion.pdf}
    \caption{{C and N abundance comparisons between different APOGEE data releases: DR14 to DR17 ({\it top panels}), and DR16 to DR17 ({\it bottom labels}). A representative error bar is shown in each panel.  The horizontal black line represents the median change. A grey shaded region is shown in each plot to represent the scatter on the median. }%\textcolor{blue}{[[SRM: Suggestions: The axes labels and tick labels in these plots are pretty small. Also, I'm not sure that I can see the grey region referred to, nor discriminate it from the horizontal line on each plot, which could be confusing?]]}
    }
    \label{C_N_abund_compare}
\end{figure*}

{The top panels of Figure \ref{C_N_abund_compare} compare the C and N abundances for star in the 75 cluster sample that have abundances in both DR14 and DR17.}
{The median change for C is measured to be $-0.029$ dex with a scatter of $0.047$ dex and the median change for N is measured to be $+0.081$ dex with a scatter of $0.052$ dex.} 
{Similarly, the bottom panels of Figure \ref{C_N_abund_compare} compare the C and N abundances for star in the 75 cluster sample that have abundances in both DR16 and DR17. The median change for C is measured to be $+0.065$ dex with a scatter of $0.039$ dex and the median change for N is measured to be $-0.013$ dex with a scatter of $0.051$ dex. These systematic offsets between the compared data releases are due to changes within the SDSS data reduction pipeline, as discussed in \S \ref{surveyDR17}}
{This simple analysis demonstrates that it is not appropriate or straightforward to inter-compare [C/N]-age relations derived from difference APOGEE data releases, unless one accounts for these simple, though systematic, variations in APOGEE carbon and nitrogen abundances over time.
}

\section{Results \& Discussion} \label{chapter3}
\subsection{Final Cluster Sample}

\begin{figure}[ht!]
\epsscale{1.18}
\plotone{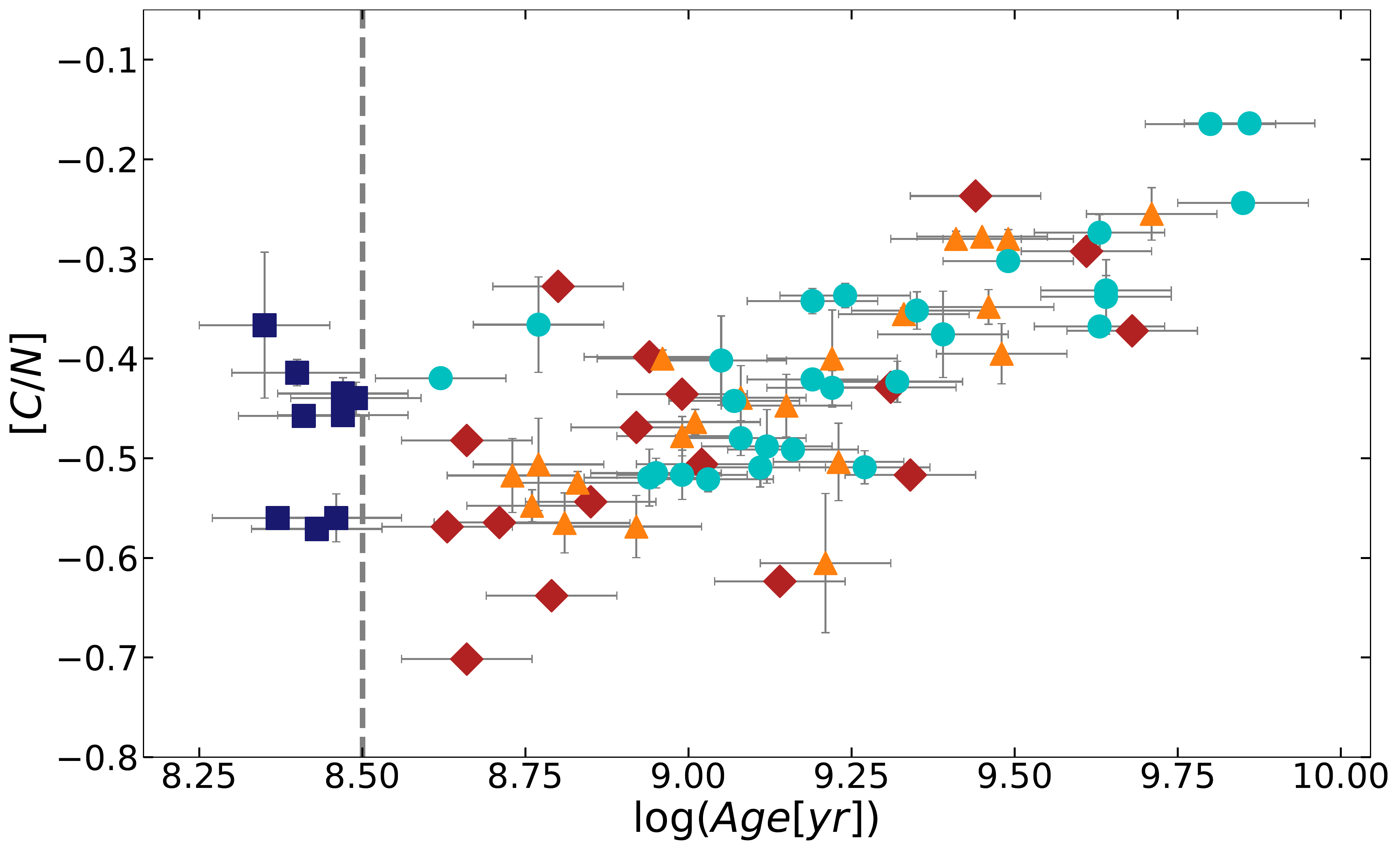}%CN_AGE_fullsample_noLinearFit_colorcode_nolegend.pdf}%{(a)}
\centering
\caption{\setlength{\baselineskip}{15.0pt} 
The [C/N] versus $\log(Age[{\rm yr}])$ distribution for clusters common to the OCCAM Survey and Cantat-Gaudin sample, which includes 75 clusters. 
Young clusters, $\log(Age) \le 8.5$, are represented with navy squares. Clusters marked with a red {diamond} are those with only one stellar member after cuts are applied. 
Clusters with 2 to 4 stellar members are represented by orange triangles, and those with 5 or more stellar members are represented by cyan circles.}
\label{fullsample}
\end{figure}

{After application of the various selection criteria summarized in Table \ref{tab:cuts}, our sample is comprised of 75 clusters with {577 member stars}. Figure \ref{fullsample} shows the [C/N] abundance ratios versus cluster age for that sample.} To ensure our calibration is reliable, we excluded clusters that contained only one member (those shown with a red {diamond}), as well as those having $\log(Age[{\rm yr}])<8.5$ (dark blue squares).
{As apparent} in Figure \ref{fullsample}, clusters younger than  $\log(Age[{\rm yr}])=8.5$ 
do not seem to follow the same trend in [C/N] and $\log(Age[{\rm yr}])$.

{{Other than possible observational issues related to the analysis of young stars \citep[][]{Baratella2020,Spina2020},} there are likely two effects at play that affect the more massive evolved stars ($M \geq 2.5M_{\odot}$) in these young clusters that are not present in the lower mass evolved stars of older clusters.}  The first is that massive stars spend less time on the first ascent RGB and proportionally more time in the core He-burning phase {\citep{Iben1967_effect1, Danchi2006}}. 
These young clusters therefore are likely dominated by stars that have already undergone a weakly degenerate or nondengeneate helium ignition and it is not yet clear whether such an event could drive additional mixing that would alter their surface abundances. 
{Secondly, in more massive stars the first dredge up reaches deeper \citep[e.g.,][]{Iben64,Iben65,Iben1967_effect1,Iben67b} into regions where the CNO cycle was significant in prior evolutionary phases. This means that the material being dredged up has a fixed ratio of carbon and nitrogen, which is at an equilibrium set by the nuclear reaction times of the CNO cycle \citep{Dearborn_1976, Dearborn_Eggleton_1976}. Since this ratio is not changing with mass, the relationship between the [C/N] ratio after the dredge up and the stellar mass becomes much weaker, making it less sensitive as an age diagnostic.}
{Given these two caveats,} we have 
excluded these stars from fits in this paper, but encourage more careful studies of stars in this regime in the future.

{After application of these additional membership and cluster age cuts the sample used for our [C/N]-age calibration (Fig.~\ref{CN_AGE_full}) is comprised of 49 clusters {(530 stars)}, covering} an age range of $8.62 \leq \log(Age[{\rm yr}]) \leq 9.82$ and a metallicity range of $-0.53 \leq {\rm [Fe/H]} \leq 0.31$. 
The individual stars from the calibration sample are listed in Table \ref{tab:full_sample_stars}.
Table \ref{sum_table} shows these calibration clusters in bold.

\subsection{The DR17 [C/N] Abundance/Age Calibration}

In log-log space the relationship between stellar age and [C/N] {appears to be linear; our best fit is given by} 
{\small
\begin{equation}
{\log[Age({\rm yr})]_{\rm DR17}} = 10.14 \, (\pm 0.08) + 2.23\,(\pm 0.19) \, {\rm [C/N]}
\label{full_cali_eqn}
\end{equation}
}
{and yields a Pearson coefficient of $R=0.79$.} Our Pearson coefficient is comparable to that found by \citet{casali_2019} {($R=0.85$)},
although we compute a slightly different offset and a weaker slope.

 \begin{figure*}[t!]
\plotone{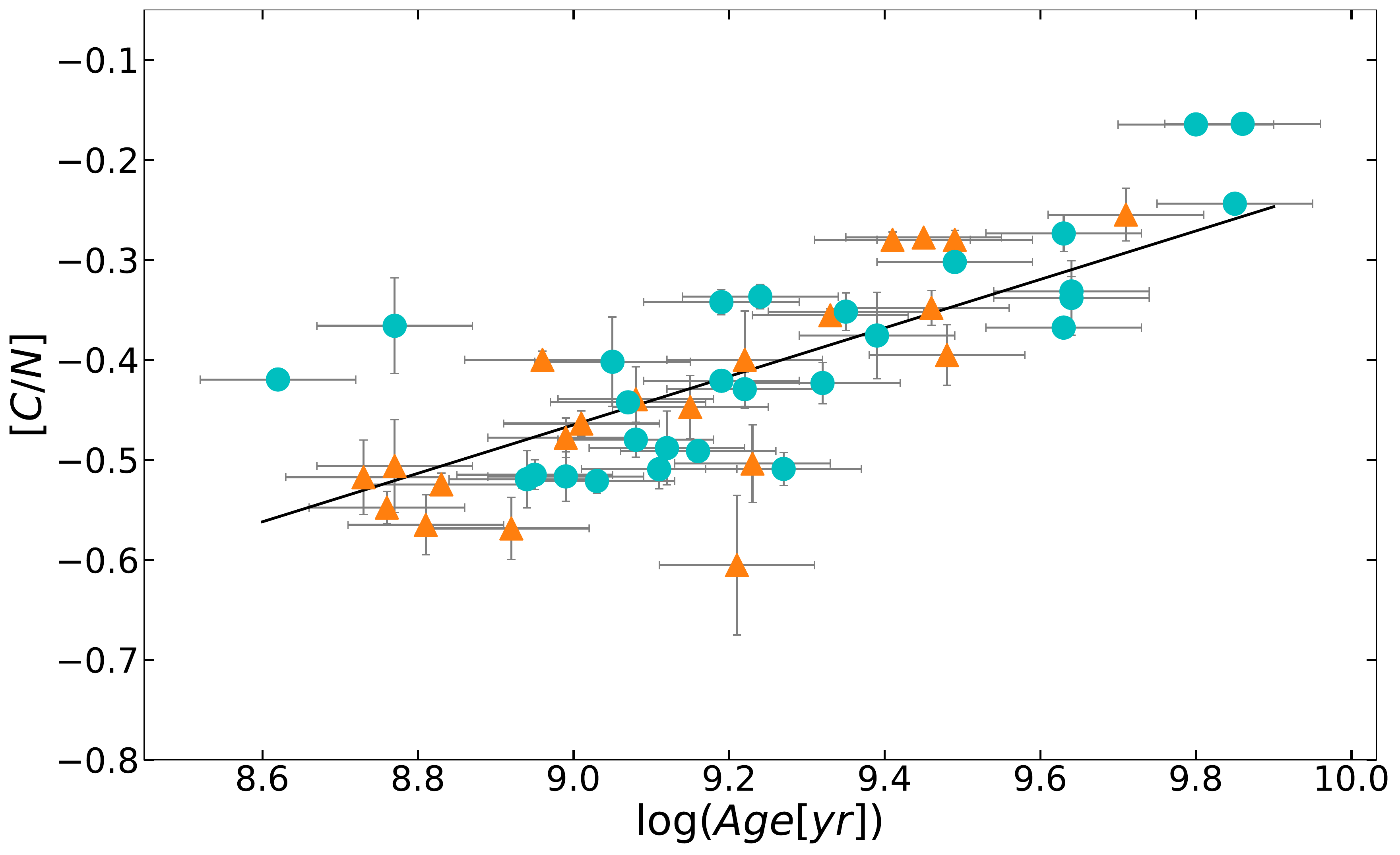}%CN_AGE_fullcalibrationsample_wLinearFit_colorcode_nolegend.pdf}%{(a)}
\centering
\caption{\setlength{\baselineskip}{15.0pt} {The [C/N] versus $\log(Age[{\rm yr}])$ distribution for the final sample, composed of clusters common to the OCCAM Survey and Cantat-Gaudin sample, totalling 49 clusters.} Clusters with 2 to 4 stellar members are represented by orange triangles, and those with 5 or more stellar members are by cyan circles. 
}
\label{CN_AGE_full}
\end{figure*}
{The linear fit given by Equation \ref{full_cali_eqn} and shown in Figure \ref{CN_AGE_full}} uses Monte Carlo (MC) resampling as described in the fits from \citet{donor_2020}. 
We computed 500 iterations of a linear fit, including uncertainties in the [C/N] abundance ratios and age, and took the mean slope and y-intercept for our final fit, where the respective errors are the standard deviation of the fit.

\begin{figure}[ht!]
\epsscale{1.18}\plotone{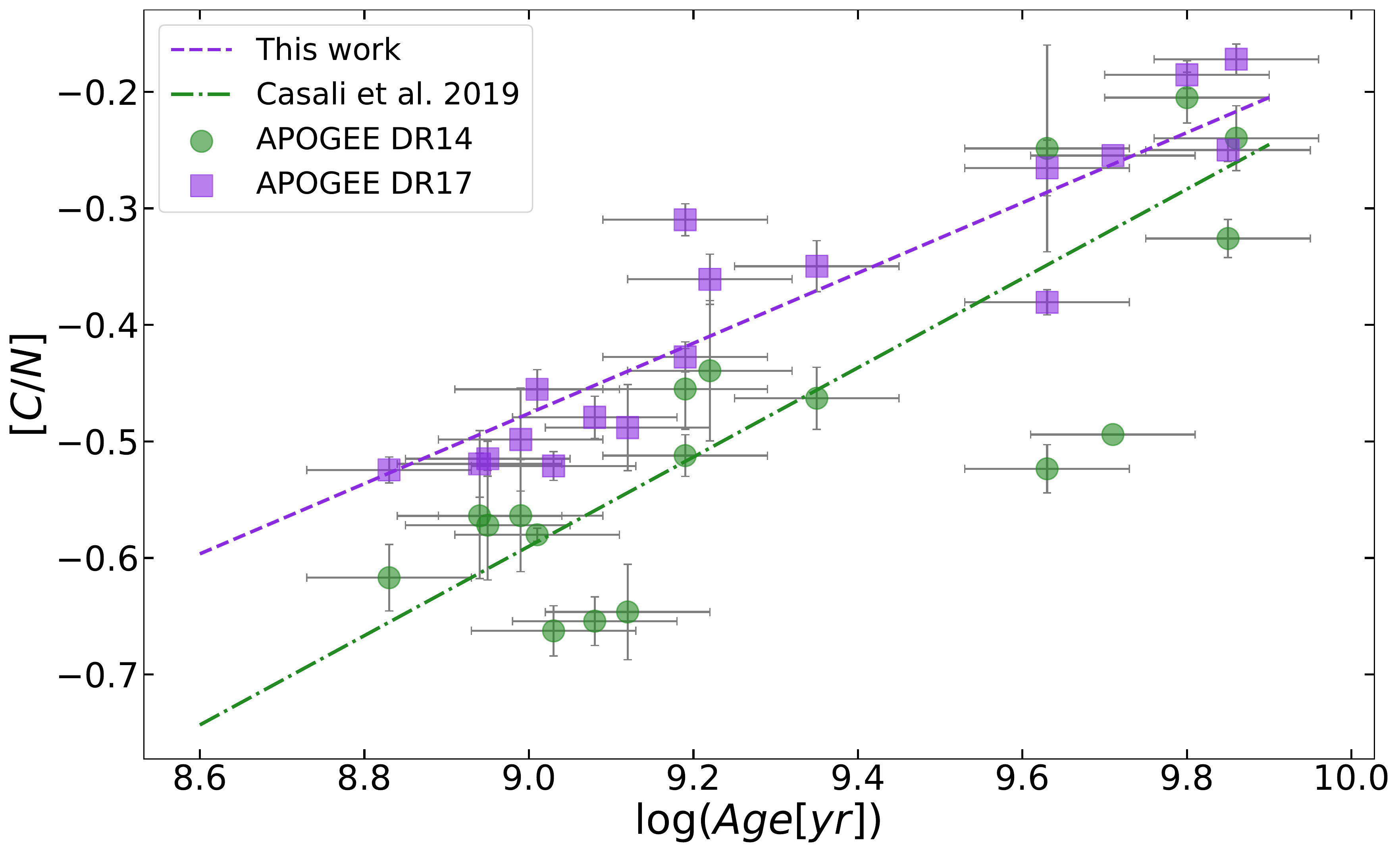}%DR14_DR17_comparison_CN_AGE_plot.pdf}%{(a)}
\centering
\caption{\setlength{\baselineskip}{15.0pt} Median [C/N] ratio of open cluster star members, as derived in DR14 (green circles) and DR17 (purple squares),
versus $\log(Ages[{\rm yr}$]), where these ages are provided by \citet{cg20}. The green dot-dashed line and purple dashed line represent the linear fit for DR14 and DR17, respectively. The plot shows
a systematic shift from DR14 data to DR17 data.}
\label{CN_AGE_DRcompare}
\end{figure}

The discrepancy between our calibration and that of \citet{casali_2019} is partially due to the different APOGEE data sets being used. \citet{casali_2019} used not only DR14 data, but also Gaia-ESO data, and this comparison is shown in Figure \ref{C_N_abund_compare}.
{\color{black}Therefore, Figure \ref{CN_AGE_DRcompare} again reveals the shift in abundance measurements between DR17 and DR14 that was used for the \citet{casali_2019} clusters.}

\subsection{Common Calibration for RC and RGB Stars?}

\begin{figure*}[]
\epsscale{1.15}\plottwo{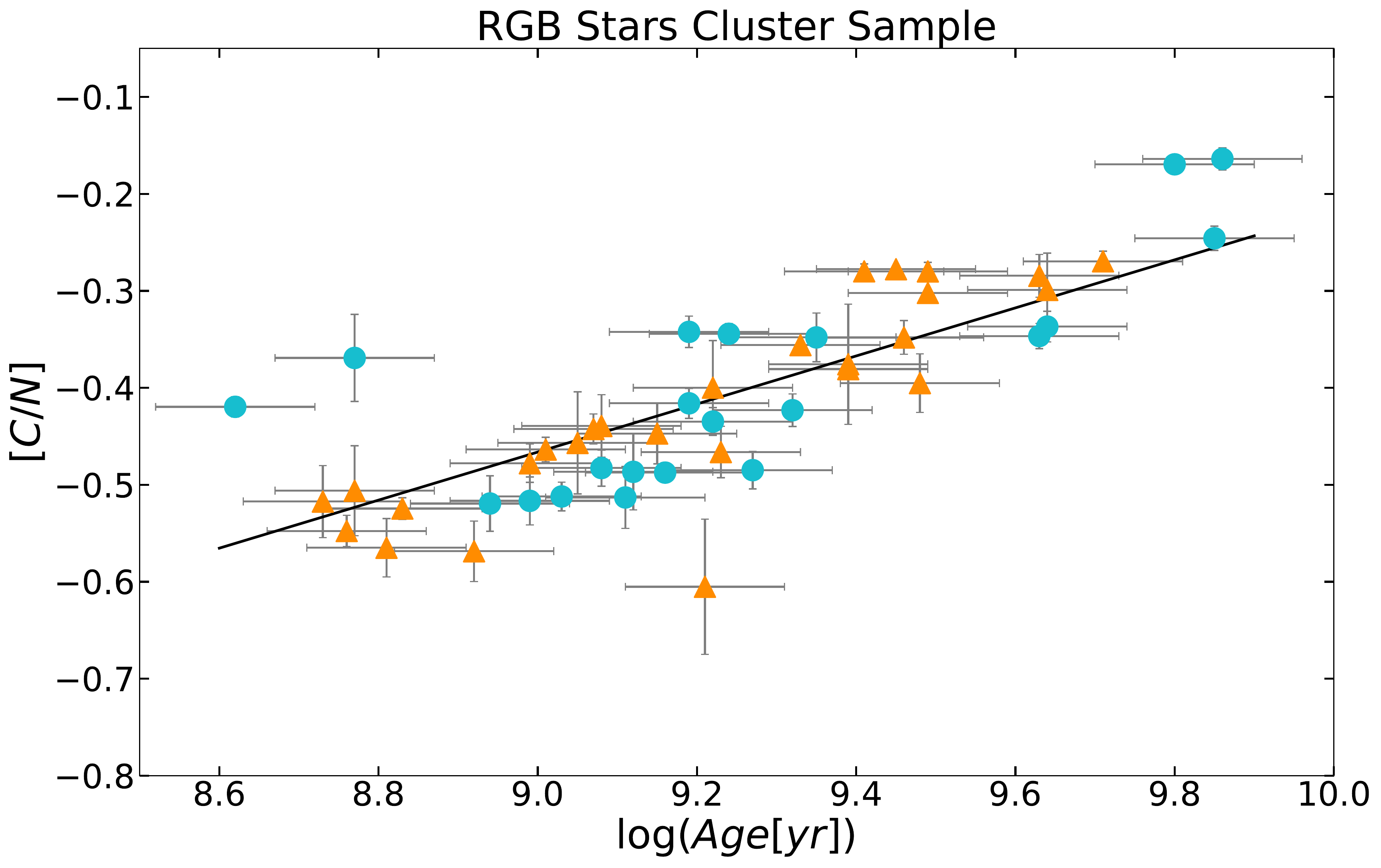}
{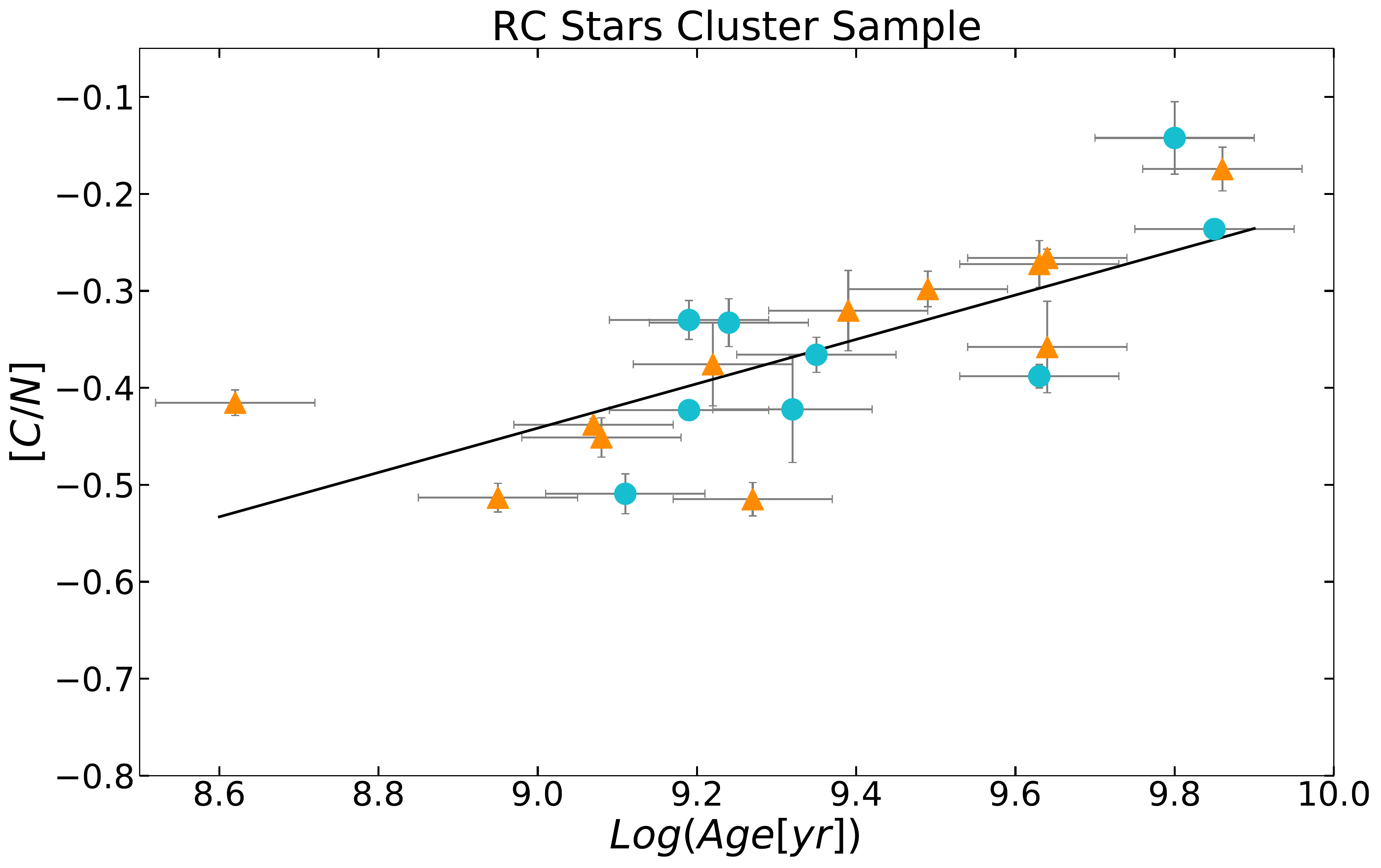}
\centering
\caption{\setlength{\baselineskip}{15.0pt} 
{({\it left}) [C/N] abundances versus log($Age[{\rm yr}$]) for our cluster sample using only RGB stars  from 46 clusters common to the OCCAM Survey and Cantat-Gaudin dataset.  ({\it right}) The [C/N] abundance versus $log(Age[{\rm yr}])$ for clusters containing only RC stars in common between the OCCAM Survey and Cantat-Gaudin dataset, which includes 21 clusters.  Labels and calculation of calibrations are similar to Figure \ref{CN_AGE_full}}
}
\label{RGB_RC_calibrations}
\end{figure*}

{Stars eventually transition} from H-burning on the RGB to He-burning on the RC.
Particularly when the He-flash is explosive, 
stars may experience significant mixing that could alter their surface abundances \citep[e.g.][]{Josiah_2020}, and therefore the relationship between [C/N] and age.
Using the APOGEE-RC catalog \citep[][Bovy et al.{\it in prep}]{bovy_2014_RC}, we have isolated RC stars from our full sample; 
once the RC stars are removed, our sample consists of 48 clusters {(400 stars)} covering an age range of $8.62 \leq \log(Age[{\rm yr}]) \leq 9.89$ and a metallicity range of $ -0.53\leq {\rm [Fe/H]} \leq 0.31$. 
A new linear fit to this sample yields
{\small
\noindent\begin{equation}
{\log[Age({\rm yr})]_{\rm DR17,RGB}} = 10.14\,(\pm 0.07) + 2.22\,(\pm 0.19)\, {\rm [C/N]}
\label{rgb_cali_eqn}
\end{equation}}
with a Pearson coefficient of 0.81, as shown in %\replaced{Figure \ref{CN_AGE_noRC}.}
{Figure \ref{RGB_RC_calibrations} ({\it left}).}
{This particular calibration yields only
a slight change in slope from the previously found trend (Eqn. \ref{full_cali_eqn}),} from $2.23(\pm 0.19)$ to $2.22(\pm 0.19)$, but more importantly, the correlation becomes more significant as the Pearson coefficient goes from $0.79$ to $0.81$.

We then created a sample comprised solely of RC stars to investigate whether RC stars follow a similar linear trend of [C/N] abundances and age to that by RGB stars. These stars yield a $log(Age)$-[C/N] relationship
that also follows a linear trend with a Pearson coefficient of 0.79:

{\small
\begin{equation}
{\log[Age({\rm yr})]_{\rm DR17,RC}} = 10.19 \, (\pm 0.09) + 2.29 \, (\pm 0.24) \, {\rm [C/N]}
\label{rc_cali_eqn}
\end{equation}
}
\noindent %\replaced{Figure \ref{CN_AGE_RConly} }
{Figure \ref{RGB_RC_calibrations} ({\it right})}
shows the relationship between the [C/N] and log(Age[Gyr]) in our RC cluster sample. The latter
is comprised of 21 clusters {(78 stars)} covering an age range of $8.62 \leq \log(Age[{\rm yr}]) \leq 9.86$ and a metallicity range of $-0.45\leq {\rm [Fe/H]} \leq 0.35$. 
We find that, within the current uncertainties, the relationship for only RGB and only RC stars are the same{. W}e therefore find no evidence for {additional} significant mixing during the He-flash that is {affecting} the surface C and N abundances {for RC stars}.

\section{Comparison to Previous Work}\label{prevwork}

{The recent [C/N]-age calibration by \citet{casali_2019} made use of APOGEE data processed with DR14 analysis software.  Moreover, for their calibration \citet{casali_2019} used both Gaia-ESO and APOGEE spectroscopic data along with a literature compilation for cluster ages. In contrast, our analysis here makes use of a uniform sample comprised of only APOGEE spectroscopic data, all reduced with the most up-to-date, DR17 reduction of those data, and the most recent \citet{cg20} uniformly determined cluster ages.}
{Thus, it would not be surprising to find significant differences between the \citet{casali_2019} relationship and that derived here.  However, surprisingly, their relationship, {$log[Age(yr)] = 10.54(\pm0.06) + 2.61(\pm0.10)[C/N]$,}
is in fair agreement with our own.}

\begin{figure*}[]
\epsscale{1.15}\plottwo{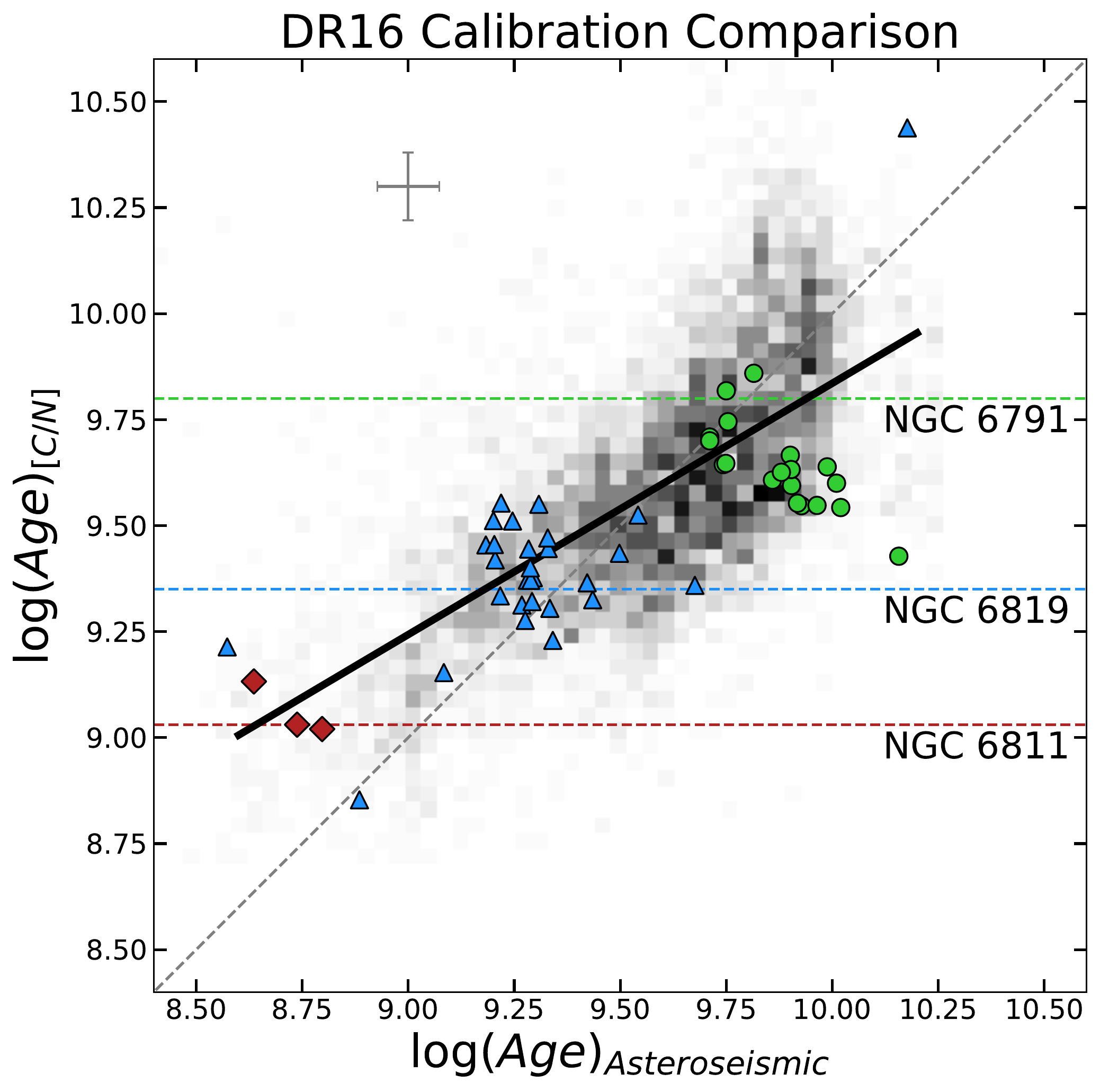}%APOKASC_DR16Calibration_AgeComparison_FULL.pdf}
{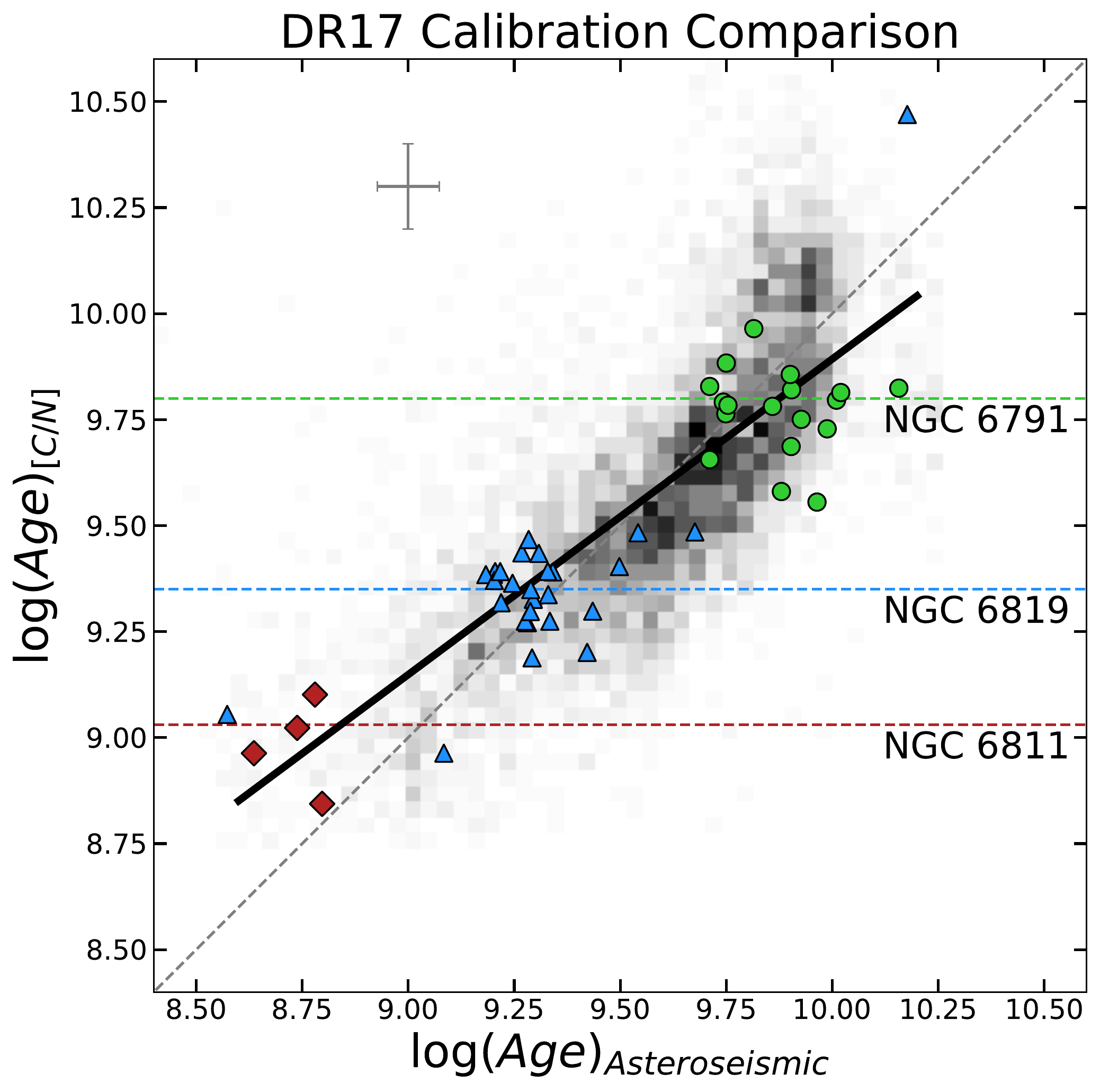}%APOKASC_DR17Calibration_AgeComparison_FULL.pdf}
\centering
\caption{\setlength{\baselineskip}{15.0pt} 
{A comparison between our DR16-based age calibration ({\it{left}}) and DR17-based age calibration ({\it{right}}) to asteroseismically determined ages from the APOKASC catalog \citep{apokasc}.} 
The grey-gradient shaded regions represent bins of field stars. 
{The green circles, blue triangles, and red diamonds represent cluster member stars common to both samples for NGC 6791, NGC 6819, and NGC 6811, respectively.}
The horizontal dashed lines are the \citet{cg20} determined ages for each cluster. The diagonal dashed grey line represents the one-to-one relationship. The solid {black} line is the linear fit to the main trend of the field stars. Mean representative error bars are shown in each panel. 
}
\label{cali_apok_age}
\end{figure*}

{As another means to verify the reliability of our methodology, we can compare predicted stellar ages from our [C/N] calibration to ages independently derived using asteroseismology.  For this purpose, we exploit the APOKASC catalog \citep[][]{apokasc}. However, because 
the chemistry used by APOKASC is based on APOGEE DR16 data, we generate} age calibrations with respect to DR16 data using the same process we used for our DR17 calibrations.Tthe calibrations for the full, RGB, and RC samples are found to be:
{\small
\begin{equation}
{\log[Age({\rm yr})]_{{\rm DR16,all}}} = 9.92 \, (\pm 0.05) + 1.94 \, (\pm 0.15)\, {\rm [C/N]}
\end{equation}
\vskip-0.26in
\begin{equation}
{\log[Age({\rm yr})]_{{\rm DR16,RGB}}} = 9.89\, (\pm 0.06) + 1.81\, (\pm 0.16)\, {\rm [C/N]}
\end{equation}
\vskip-0.26in
\begin{equation}
{\log[Age({\rm yr})]_{{\rm DR16,RC}}} = 9.88\, (\pm 0.07) + 1.94\, (\pm 0.25)\, {\rm [C/N]}
\end{equation}
}
\noindent We apply these DR16 full, RGB, and RC sample calibrations and the DR17 full, RGB, and RC sample calibrations (Eqns. \ref{full_cali_eqn}, \ref{rgb_cali_eqn}, and \ref{rc_cali_eqn}, respectively) on field stars within the APOKASC sample and compare the [C/N]-based stellar ages using our calibrations to the astroseismology-based stellar ages determined by APOKASC.
We find the [C/N] based ages are more consistent with the asteroseismic ages when the newer DR17 calibration is used (see Figure \ref{cali_apok_age}), which suggests that the inferred [C/N] values have improved with time and that the newer DR17 data should be preferred.

For further verification, we apply our calibrations to open clusters that are also in the APOKASC sample: NGC 6791, NGC 6819, and NGC 6811. We find that our calculated ages from the DR17 calibration are closer to the ages determined in \citet{cg20} for all three clusters. The spread in the [C/N]-calibration ages is due to the uncertainty of [C/N] in the individual cluster members, but such a spread is expected because our calibration is based on the average [C/N] abundance of the cluster.
Comparison with these asteroseismic results suggests that [C/N] based ages can be trusted to 10\% for 99.79\% of giants, as shown in Figure \ref{cali_apok_age}.  

On the upper giant branch, above {the so-called ``red bump'',} extra convective mixing can occur, particularly in low-metallicity {([Fe/H]$ \lesssim -0.5$)} stars \citep{Gratton_2000, Shetrone_2019}. {In principle, this would significantly impact} the ages inferred for these stars, making them appear to be younger than they actually are. For this reason, we caution again extrapolating our results too far outside the calibration metallicity range {$-0.53 \lesssim {\rm [Fe/H]} \lesssim 0.31$},
particularly for stars on the upper red giant branch.

\clearpage
\section{Conclusions}
{The [C/N] age-dating technique, applicable to RGB stars, has brought a powerful and versatile tool that can be applied to huge numbers of Galactic field stars explored by large-scale spectroscopic surveys with which these two chemical species can be measured, such as APOGEE DR17.}
In this work, we have used open cluster ages to calibrate [C/N] as a chemical clock for evolved stars, specifically for the SDSS/APOGEE survey.

\begin{enumerate}
\item This calibration, based on APOGEE DR17, provides the following relation 
for all evolved stars that experience the first dredge-up (e.g., APOGEE DR17 stars with $\log g < 3.3$):
\noindent\begin{equation}
{\log[{\it Age}({\rm yr})]} = 2.23 \, (\pm 0.19)\, {\rm [C/N]} + 10.14 \, (\pm 0.08) 
\end{equation}
\noindent {This calibration is found to be independent of metallicity of the range explored, $-0.5\leq[Fe/H]\leq+0.3$ with our sample. 
The calibration is consistent, though slightly different, with the [C/N]-age calibration using DR14 and Gaia-ESO by \citet{casali_2019}, which is primarily due to differences in the C and N abundance determination between DR14 and DR17.
We also determined the calibration for DR16 which is presented in \S \ref{prevwork}.}

\item We show that the relationship between age and [C/N] for massive ($M \geq 2.5 M_{\odot}$) young {(log$(Age)<8.5$)} giants does not follow the same trend as the relationship calibrated for older, lower mass giants here. We therefore caution that our relationship should not be {extrapolated} to this regime. 

\item We find that RC stars can be fit with the same calibration as RGB stars. Therefore, we see no evidence for significant extra mixing that affects C and N on the upper RGB or during the He-flash for low-mass metal rich giants. 
{These results align with findings of \citet{Shetrone_2019}. With this in mind, we caution the reader not to extrapolate the calibration to lower metallicities ([Fe/H] $\lesssim -0.5$), where potential extra-mixing might have a significant impact in measured elemental abundances.} 

\item {\color{black} Comparison with asteroseismic results suggests that [C/N] based ages can be trusted to 10\% for 99.79\% of giants, 
opening up an exciting future for the estimation of {\it precise} and {\it accurate} ages for hundreds of thousands of {evolved} stars across the Galaxy.} 

\end{enumerate}

%\clearpage
\begin{acknowledgements}
TS, PMF, NM, and JD acknowledge support for this research from the National Science Foundation (AST-1715662). JT and PMF acknowledge this work was performed at the Aspen Center for Physics, which is supported by National Science Foundation grant PHY-1607611.
JT acknowledges support for this work was provided by NASA through the NASA Hubble Fellowship grant No.51424 awarded by the Space Telescope Science Institute, which is operated by the Association of Universities for Research in Astronomy, Inc., for NASA, under contract NAS5-26555 was supported in part by the National Science Foundation under Grant No. NSF PHY-1748958.
{KC acknowledges support for this research from the
National Science Foundation (AST-0907873).}
%\textcolor{red}{
{SRM acknowledges support from National Science Foundation award AST-1908331.}
{DAGH acknowledges support from the State Research Agency (AEI) of the
Spanish Ministry of Science, Innovation and Universities (MCIU) and the
European Regional Development Fund (FEDER) under grant AYA2017-88254-P.}
\end{acknowledgements}

Funding for the Sloan Digital Sky 
Survey IV has been provided by the 
Alfred P. Sloan Foundation, the U.S. 
Department of Energy Office of 
Science, and the Participating 
Institutions. 

SDSS-IV acknowledges support and 
resources from the Center for High 
Performance Computing  at the 
University of Utah. The SDSS 
website is www.sdss.org.

SDSS-IV is managed by the 
Astrophysical Research Consortium 
for the Participating Institutions 
of the SDSS Collaboration including 
the Brazilian Participation Group, 
the Carnegie Institution for Science, 
Carnegie Mellon University, Center for 
Astrophysics | Harvard \& 
Smithsonian, the Chilean Participation 
Group, the French Participation Group, 
Instituto de Astrof\'isica de 
Canarias, The Johns Hopkins 
University, Kavli Institute for the 
Physics and Mathematics of the 
Universe (IPMU) / University of 
Tokyo, the Korean Participation Group, 
Lawrence Berkeley National Laboratory, 
Leibniz Institut f\"ur Astrophysik 
Potsdam (AIP),  Max-Planck-Institut 
f\"ur Astronomie (MPIA Heidelberg), 
Max-Planck-Institut f\"ur 
Astrophysik (MPA Garching), 
Max-Planck-Institut f\"ur 
Extraterrestrische Physik (MPE), 
National Astronomical Observatories of 
China, New Mexico State University, 
New York University, University of 
Notre Dame, Observat\'ario 
Nacional / MCTI, The Ohio State 
University, Pennsylvania State 
University, Shanghai 
Astronomical Observatory, United 
Kingdom Participation Group, 
Universidad Nacional Aut\'onoma 
de M\'exico, University of Arizona, 
University of Colorado Boulder, 
University of Oxford, University of 
Portsmouth, University of Utah, 
University of Virginia, University 
of Washington, University of 
Wisconsin, Vanderbilt University, 
and Yale University.

This work has made use of data from the European Space Agency (ESA) mission {\it Gaia} (\url{https://www.cosmos.esa.int/gaia}), processed by the {\it Gaia} Data Processing and Analysis Consortium (DPAC, \url{https://www.cosmos.esa.int/web/gaia/dpac/consortium}). Funding for the DPAC has been provided by national institutions, in particular the institutions participating in the {\it Gaia} Multilateral Agreement.

%This research made use of Astropy, a community-developed core Python package for Astronomy (Astropy Collaboration, 2018).
This research made use of Astropy, a community-developed core Python package
    for Astronomy \citep{astropy:2013,astropy:2018}.

% \end{acknowledgements}

\facilities{Du Pont (APOGEE), Sloan (APOGEE), Spitzer, WISE, 2MASS, Gaia}
\software{\href{http://www.astropy.org/}{Astropy}~\citep{astropy:2013,astropy:2018}} 
%\clearpage
\bibliography{Spoo}{}
\bibliographystyle{aasjournal}

\listofchanges

\end{document}